\documentclass[12pt]{iopart}
\usepackage{amssymb} 
\usepackage[dvips]{graphicx}

\begin{document}

\title[On the influence of statistics on the determination of $\langle X_{max} \rangle$]{On the influence of 
statistics on the determination of the mean value of the depth of shower maximum for ultra high energy cosmic 
ray showers}

\author{A.~D. Supanitsky}
\address{Instituto de Astronom\'ia y F\'isica del Espacio (IAFE), UBA-CONICET, Argentina.}
\ead{supanitsky@iafe.uba.ar}
\author{G. Medina-Tanco}
\address{Instituto de Ciencias Nucleares, UNAM, Circuito Exteriror S/N, Ciudad Universitaria, M\'exico 
D. F. 04510, M\'exico.}
\ead{gmtanco@nucleares.unam.mx}

\begin{abstract}
The chemical composition of ultra high energy cosmic rays is still uncertain. The latest results obtained by 
the Pierre Auger Observatory and the HiRes Collaboration, concerning the measurement of the mean value and 
the fluctuations of the atmospheric depth at which the showers reach the maximum development, $X_{max}$, are 
inconsistent. From comparison with air shower simulations it can be seen that, while the Auger data may be 
interpreted as a gradual transition to heavy nuclei for energies larger than $\sim 2-3\times10^{18}$ eV, the 
HiRes data are consistent with a composition dominated by protons. In Ref. \cite{Wilk:11} it is suggested that 
a possible explanation of the observed deviation of the mean value of $X_{max}$ from the proton expectation, 
observed by Auger, could originate in a statistical bias arising from the approximated exponential shape of the 
$X_{max}$ distribution, combined with the decrease of the number of events as a function of primary energy. In 
this paper we consider a better description of the $X_{max}$ distribution and show that the possible bias in 
the Auger data is at least one order of magnitude smaller than the one obtained when assuming an exponential 
distribution. Therefore, we conclude that the deviation of the Auger data from the proton expectation is 
unlikely explained by such statistical effect.    
\end{abstract}

\maketitle

\section{Introduction}

The nature of the primary cosmic rays is intimately related to the astrophysical objects capable
of accelerating these particles to such high energies. Also, propagation in the intergalactic medium
depends on the composition, which affects the resulting spectral distribution of the flux observed at 
Earth. A knowledge of the composition is also very important for primary energy reconstruction and 
for anisotropy studies.   
 
One of the most important limitations of composition analyses comes from the lack of knowledge of the 
hadronic interactions at the highest energies. Composition studies are based on the comparison of 
experimental data with Monte Carlo simulations of atmospheric cosmic rays showers, which makes use of 
hadronic interaction models which extrapolate the available low energy accelerator data to the energies 
of the cosmic rays.    

One of the most sensitive parameters to the mass of the primary cosmic ray is the atmospheric depth 
at which the showers reach their maximum development. Lighter primaries generate showers that are more 
penetrating, producing larger values of $X_{max}$. Also, the fluctuations of this parameter are smaller 
for heavier nuclei. The Pierre Auger Observatory and the HiRes experiment are able to observe directly
the longitudinal development of the showers by means of fluorescence telescopes. Therefore, in both 
experiments, the $X_{max}$ parameter of each observed shower can be reconstructed from 
the data taken by the telescopes. 

The mean value and the standard deviation of $X_{max}$, as a function of primary energy, obtained by 
Auger \cite{Auger:10} and HiRes \cite{Hires:10} appear to be inconsistent. From the comparison with 
simulations, the Auger data suggest a transition to heavier nuclei starting at energies of order of 
$2-3\times10^{18}$ eV, whereas, the HiRes data are consistent with protons in the same energy range. 
In Ref. \cite{Wilk:11} a new parameter, the difference between the mean value and the standard deviation 
of $X_{max}$, was introduced in order to reconcile the Auger and HiRes results. This new parameter has 
the advantage of being much less sensitive to the first interaction point than the mean value and the 
standard deviation separately. From a comparison of the experimental values of this parameter, obtained 
by Auger and HiRes, with simulated data, they infer that the composition of the cosmic rays is dominated 
by protons. They say that the energy dependence of the distribution of $X_{max}$, observed by Auger, seems 
to be caused by an unexpected change in the depth of the first interaction point, which can be explained 
by a rapid increase of the cross section and/or increase of the inelasticity. Both possibilities require 
an abrupt onset of new physics in this energy range, which makes them questionable. They also suggest 
that the deviation of the distribution of $X_{max}$ from the proton expectation, present in the Auger 
data, could be originated in the statistical techniques used to analyze the data. In particular, they suggest 
that the deviation of the mean value of $X_{max}$ from the proton expectation could be explained by a bias 
originated from the exponential nature of the $X_{max}$ distribution and the decreasing number of events 
as a function of primary energy.   
 
In this work we show that, considering a better description of the $X_{max}$ distribution, the bias in 
the determination of the mean value of $X_{max}$ become more than one order of magnitude smaller than 
the one obtained for the exponential distribution. We find that the value of the bias in the last energy 
bin (the one with the smallest number of events) of the Auger data, published in Ref. \cite{Auger:10}, 
is $\lesssim 1.5$ g cm$^{-2}$, which is much smaller than the systematic errors on the determination of 
the mean value of $X_{max}$ estimated in Ref. \cite{Auger:10}.

\section{Numerical approach}

Following Ref. \cite{Wilk:11} let us introduce the parameter,
\begin{equation}
\xi(N)=1-\frac{\textrm{mode}[\bar{X}_{max}^N]}{\langle X_{max} \rangle},
\label{xi}
\end{equation}
where $\langle X_{max} \rangle$ is the mean value of the $X_{max}$ distribution,
\begin{equation}
\label{SMean}
\bar{X}_{max}^N = \frac{1}{N} \sum_{i=1}^{N} X_{max}^i,
\end{equation}
is the sample mean corresponding to samples of size $N$ and $\textrm{mode}[\bar{X}_{max}^N]$ is the 
value of $\bar{X}_{max}^N$ that occurs most frequently, i.e. the maximum of the distribution function 
of $\bar{X}_{max}^N$. Therefore, the bias on the determination of $\langle X_{max} \rangle$ appears when
a particular realization of the sample mean is equal to the mode of the sample mean distribution function. 
Note that the sample mean (Eq. (\ref{SMean})) is an unbiased estimator of the mean of the exponential 
distribution, i.e. $E[\bar{X}_{max}^N]=\langle X_{max} \rangle$. In Ref. \cite{Wilk:11} it is shown that 
approximating the $X_{max}$ distribution by an Exponential function the parameter $\xi(N)$ is given by: 
$\xi_E(N)=1/N$.  

In order to better describe the distribution of $X_{max}$ two different types of functions are considered. They are 
chosen in such a way that the distribution of $\bar{X}_{max}^N$ can be obtained, at least, in a semi-analytical 
way. The first function considered is a shifted-Gamma distribution \cite{Schmidt:08},
\begin{equation}
\label{Gamma}
P_{G}(X_{max})=\left\{ 
\begin{array}{ll}
  \mathop{\displaystyle \frac{(X_{max}-X_{0})^{k-1}}{\Gamma(k)\ \tau_X^k} \exp \left( -\frac{X_{max}-X_{0}}{\tau_X} \right) } &  X_{max} \geq X_{0} \\
  0                                 &  X_{max} < X_{0}
\end{array}  \right.,
\end{equation}
where $k=5$ and the other two parameters can be obtained from the mean value and the standard deviation of 
$X_{max}$,
\begin{eqnarray}
\label{X0}
X_{0} &=& \langle X_{max} \rangle-k\ \tau_X, \\
\label{tau}
\tau_X &=& \frac{\sigma[X_{max}]}{\sqrt{k}}.
\end{eqnarray}

The second function under consideration is the convolution between an exponential function and a Gaussian (Exp-Gauss),
\begin{eqnarray}
\label{ExpGauss}
P_{EG}(X_{max})&=& \frac{1}{\lambda \sqrt{2 \pi} \beta} \int_{-\infty}^{X_{max}} du\ \exp \! \left( -\frac{X_{max}-u}{\lambda}  \right)%
\exp \! \left(-\frac{(u-\alpha)^2}{2 \beta^2}  \right) \nonumber \\
&=& \frac{1}{2 \lambda} \exp \! \left( -\frac{X_{max}-\alpha}{\lambda} +\frac{\beta^2}{2 \lambda^2} \right) %
\textrm{Erfc} \! \left(\frac{\beta}{\sqrt{2} \lambda}-\frac{X_{max}-\alpha}{\sqrt{2} \beta} \right) \! ,
\end{eqnarray}
where $\alpha,\ \beta$ and $\lambda$ are fitting parameters and 
\begin{equation}
\textrm{Erfc}(z)=1-\frac{2}{\sqrt{\pi}} \int_{0}^{z} dt\ \exp\left( -t^2/2  \right).
\end{equation}

A library of simulated showers was generated by using the program CONEX (v2r2.3) \cite{conex}. Monochromatic samples 
of $10^4$ proton showers were generated from $\log(E/eV)=18$ to $\log(E/eV)=19.5$ in steps of $\Delta \log(E/eV)=0.1$. 
The arrival directions of the showers follow an isotropic distribution, such that the zenith angle is in the interval 
$[0^\circ,\ 60^\circ]$. The hadronic interaction models considered are QGSJET-II \cite{QGSJETII} and EPOS 1.99 \cite{Epos}. 

\begin{figure}[!h]
\centering
\includegraphics[width=7.5cm]{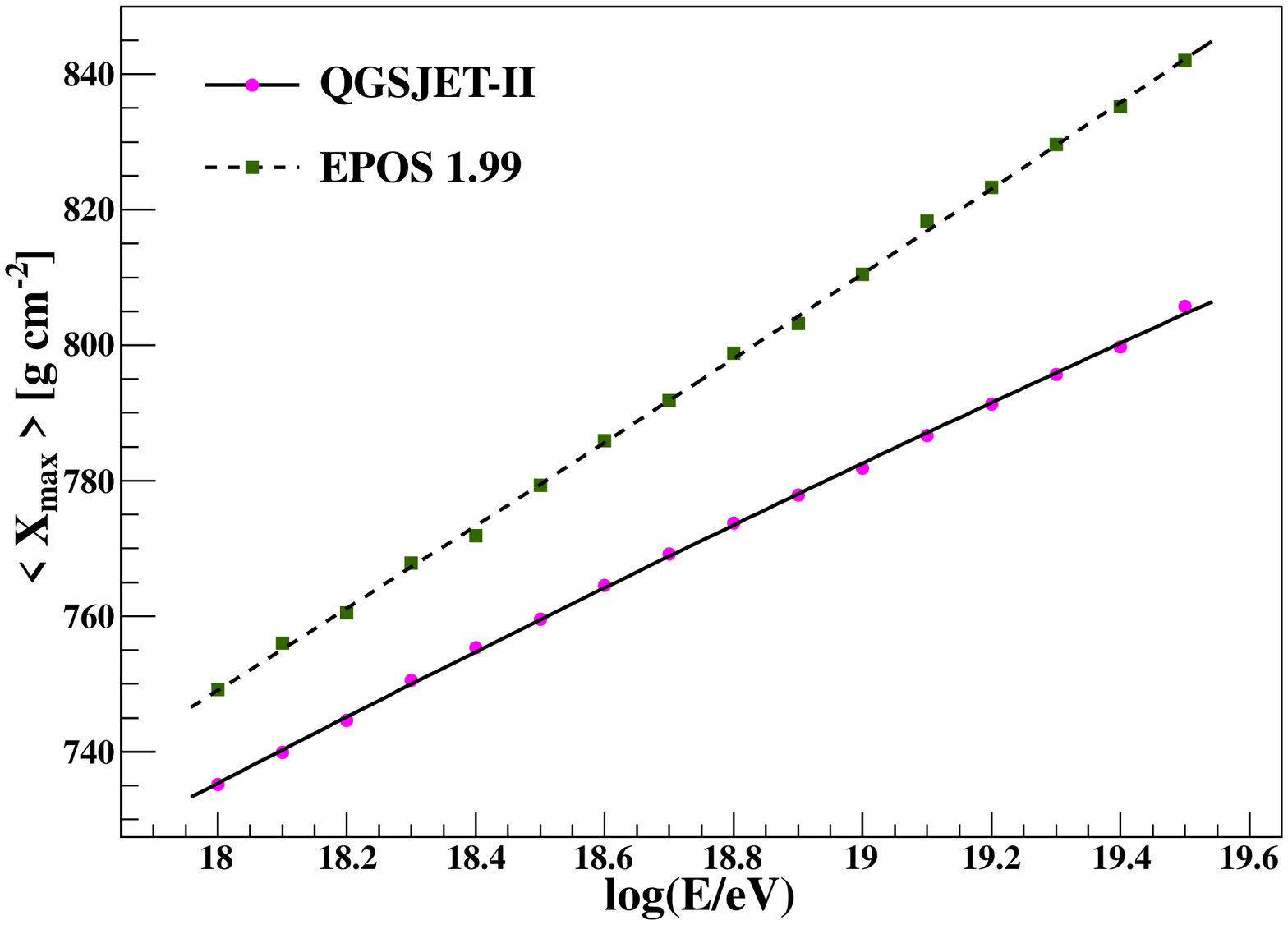}
\includegraphics[width=7.5cm]{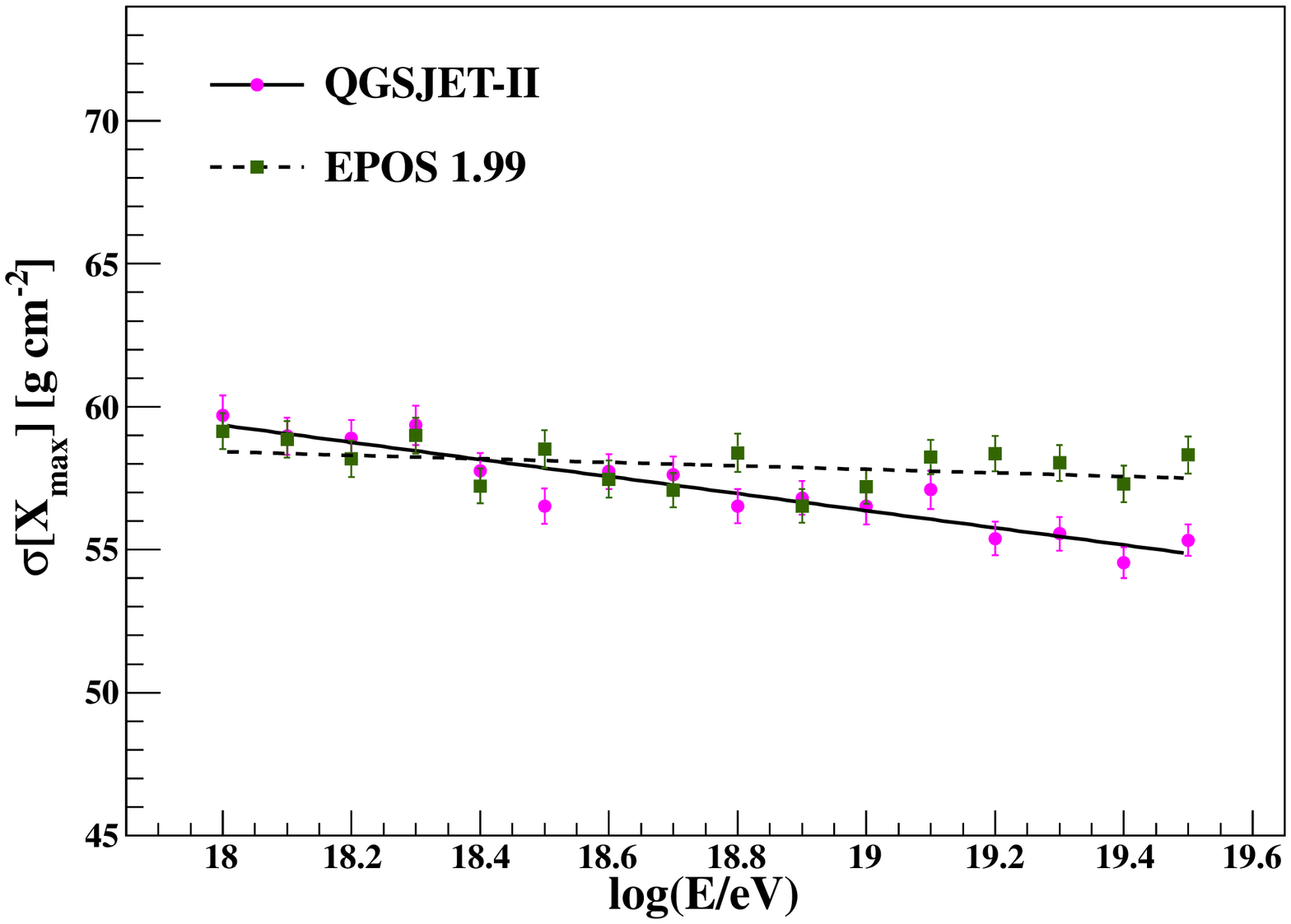}
\caption{Mean value (left panel) and the standard deviation (right panel) of $X_{max}$ as a function of $\log(E/eV)$ 
obtained by using CONEX with QGSJET-II and EPOS 1.99 for proton initiated showers. The lines correspond to the fits
of the simulated data (see the text for details).}
\label{MSFits}
\end{figure}

The mean value and the standard deviation (needed for the description of the $X_{max}$ distribution using the 
shifted-Gamma function) were fitted with a quadratic function and a linear function of $\log(E)$, respectively, i.e.,
\begin{eqnarray}
\label{mean}
\langle X_{max} \rangle &=& A_0+A_1 \log(E/eV) + A_2 \log^2(E/eV), \\
\label{sig}
\sigma[X_{max}] &=& B_0+B_1 \log(E/eV).
\end{eqnarray} 
Figure \ref{MSFits} shows the simulated data as well as the fits, for both hadronic interaction models considered.
The values of the parameters corresponding to Eqs. (\ref{mean}) and (\ref{sig}) are given in table \ref{param}.
\begin{table}[!h]
\caption{\label{param} Parameters corresponding to the quadratic and linear fits of $\langle X_{max} \rangle$
and $\sigma[X_{max}]$, respectively (see Eqs. (\ref{mean}) and (\ref{sig})), obtained from simulations for QGSJET-II 
and EPOS 1.99.}
\begin{tabular*}{\textwidth}{@{}l*{15}{@{\extracolsep{0pt plus12pt}}l}}
\br
& $A_0$ [g cm$^{-2}$] & $A_1$ [g cm$^{-2}$] & $A_2$ [g cm$^{-2}$] & $B_0$ [g cm$^{-2}$] & $B_1$ [g cm$^{-2}$] \\
\mr
QGSJET-II & -826.171 & 124.198 & -2.08037 & 113.223 & -2.99273  \\
EPOS 1.99 & 80.4419  & 14.1183 & 1.27933  & 69.4862 & -0.614616 \\
\br
\end{tabular*}
\end{table}

The distribution functions of $X_{max}$, for every energy and hadronic interaction model considered, were fitted 
with the Exp-Gauss function, Eq. (\ref{ExpGauss}). The parameters $\alpha, \ \beta$ and $\lambda$ were fitted with 
linear functions of $\log(E)$, in order to obtain the Exp-Gauss representation of the $X_{max}$ distribution for 
every value of energy in the interval $[10^{18},\ 10^{19.5}]$ eV, see \ref{app1} for details. 

\begin{figure}[!h]
\centering
\includegraphics[width=7.5cm]{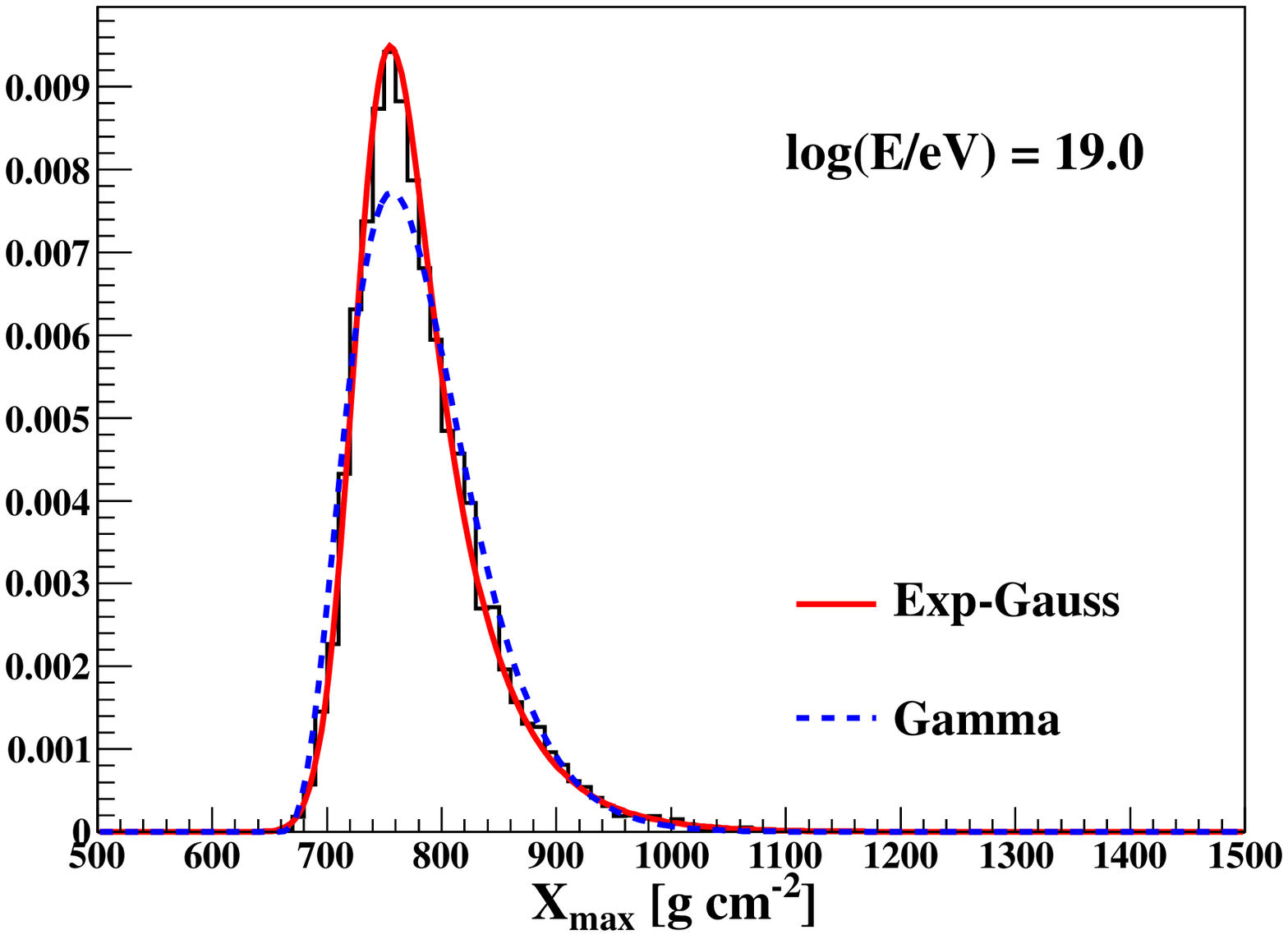}
\includegraphics[width=7.5cm]{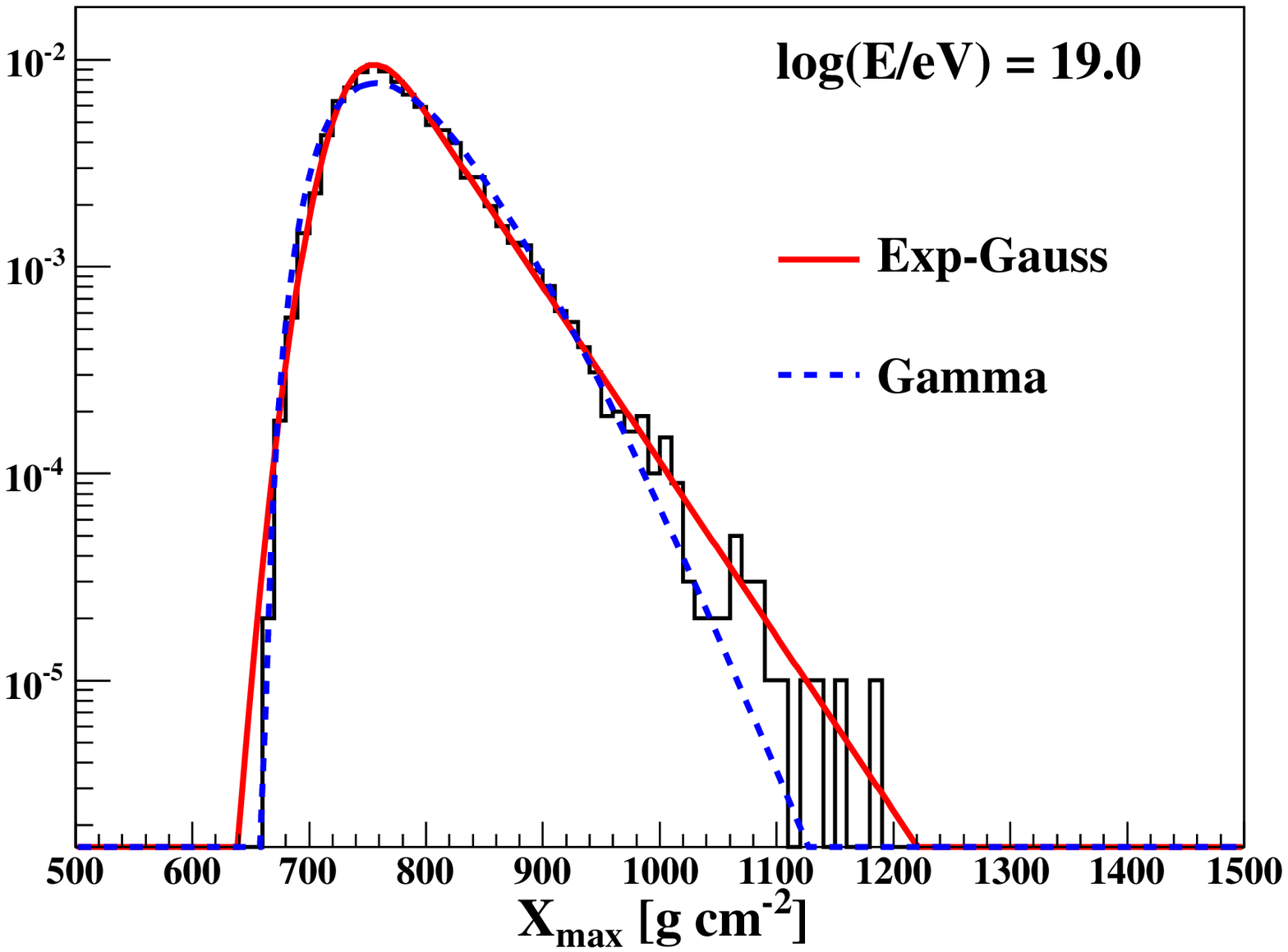}
\includegraphics[width=7.5cm]{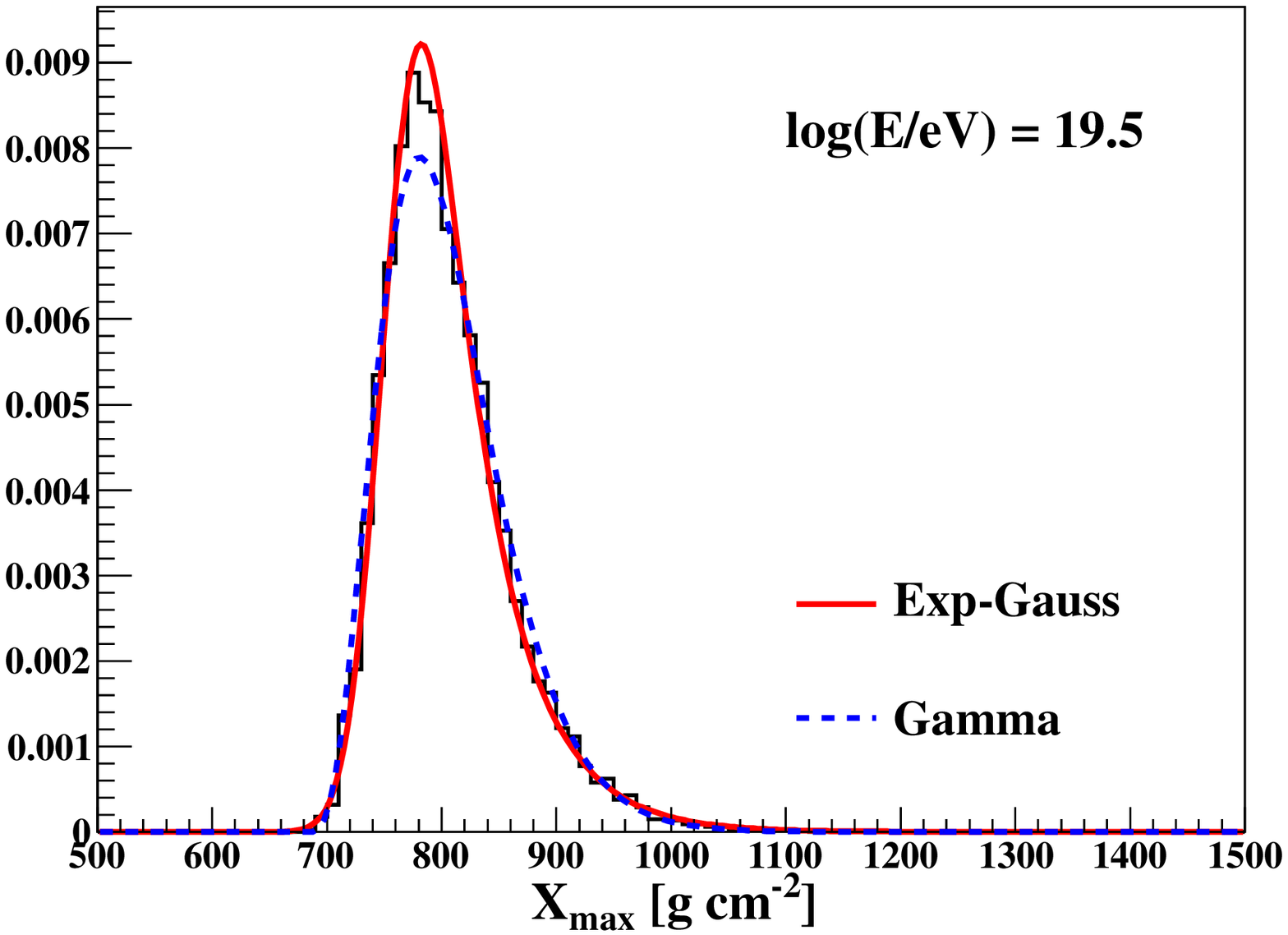}
\includegraphics[width=7.5cm]{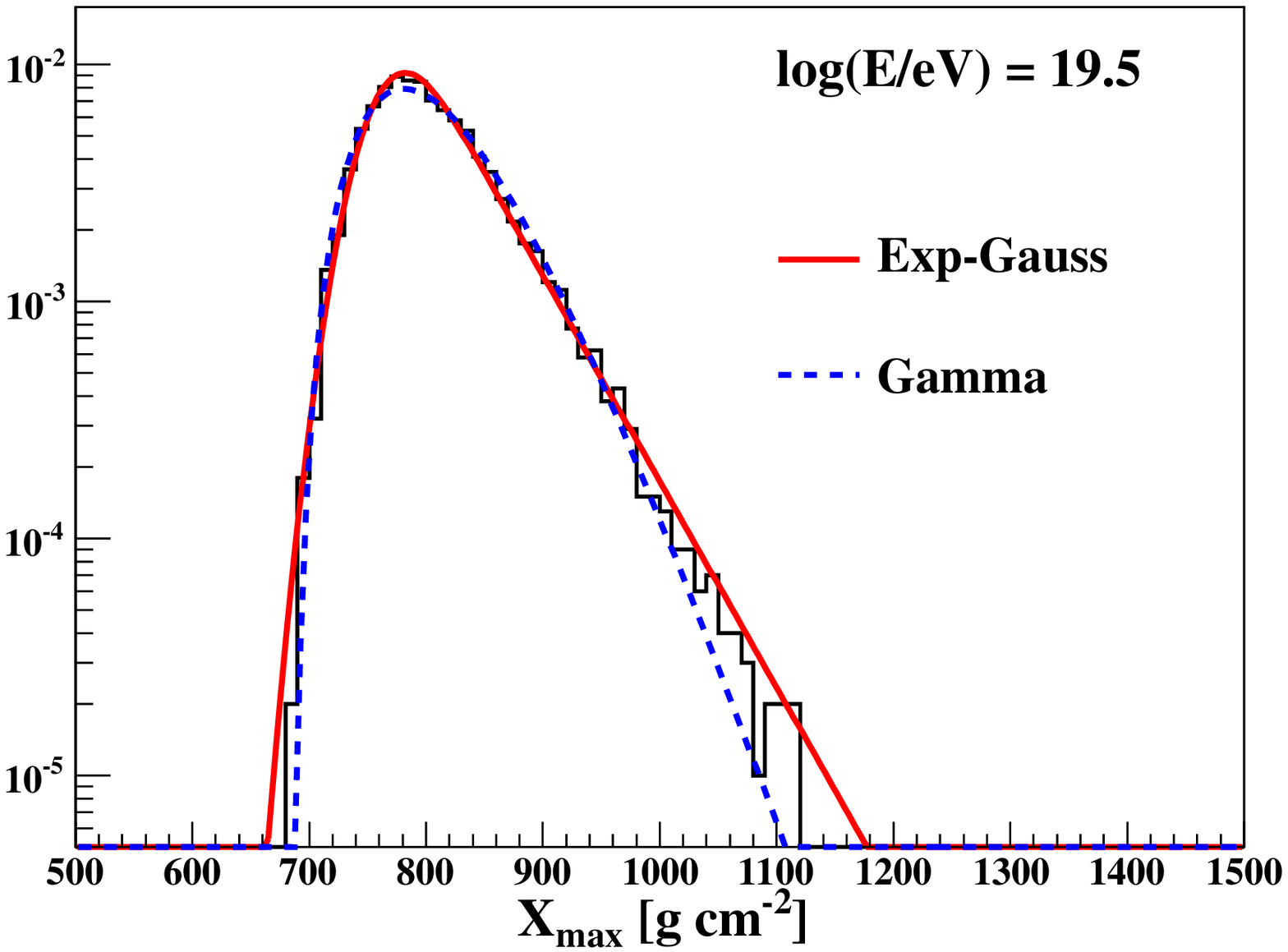}
\caption{Distributions of $X_{max}$ for proton showers generated by using CONEX with QGSJET-II. Red solid lines 
correspond to the fits of the histograms with the Exp-Gauss function, Eq. (\ref{ExpGauss}). The blue dashed lines 
correspond to the shifted-Gamma function, Eq. (\ref{Gamma}), where the parameters $X_0$ and $\tau_X$ are obtained 
by using Eqs. (\ref{X0}), (\ref{tau}), and the fits of $\langle X_{max} \rangle$ and $\sigma[X_{max}]$ as a function 
of $\log E$ (see text for details).}
\label{XmaxFits}
\end{figure}

Figure \ref{XmaxFits} shows the distributions of $X_{max}$, obtained by using CONEX with QJSJET-II, for 
$\log(E/eV)=19$ and $\log(E/eV)=19.5$. Red solid lines correspond to the fits of the simulated data with the 
Exp-Gauss function. The blue dashed lines correspond to the shifted-Gamma function, Eq. (\ref{Gamma}), 
for which the parameters $X_0$ and $\tau_X$ are obtained by using the expressions of $\langle X_{max} \rangle$ 
and $\sigma[X_{max}]$ in Eqs. (\ref{mean}) and (\ref{sig}) to calculate $X_0$ and $\tau_X$ from Eqs. (\ref{X0}) 
and (\ref{tau}), respectively. From the figure it can be seen that the Exp-Gauss function is a better fit to the 
simulated data than the shifted-Gamma function. It can also be seen that the tail to larger values of $X_{max}$ is 
slightly overestimated by the Exp-Gauss distribution and underestimated by the shifted-Gamma function. Therefore, 
the distribution function of the universe (samples with $N \rightarrow \infty$) should fall between this two 
functions.     

The distribution of $\bar{X}_{max}^N$ can be calculated by means of the characteristic function, which is defined as 
the expectation value of $\exp(i t X_{max})$, i.e. $\phi_{X_{max}}(t) = E[\exp(i t X_{max})]$. It is straightforward 
to show that the characteristic function of $\bar{X}_{max}^N$ is given by $\phi_{\bar{X}_{max}^N}(t) = [\phi_{X_{max}}(t/N)]^N$ 
\cite{James:08}. 

The characteristic function of the shifted-Gamma distribution is $\phi_{X_{max}}^G(t) = \exp(i X_0 t)\ (1-i t \tau_X)^{-k}$ 
and then the characteristic function of $\bar{X}_{max}^N$ is given by 
$\phi_{\bar{X}_{max}^N}^G(t) = \exp(i X_0 t)\ (1-i t \tau_X/N)^{-k N}$, which corresponds also to a shifted-Gamma distribution. 
Therefore, the distribution function of $\bar{X}_{max}^N$ is given by, 
\begin{equation}
\label{AvGamma}
\bar{P}_{G}(\bar{X}_{max}^N)=\left\{ 
\begin{array}{ll}
  \mathop{\displaystyle \frac{(\bar{X}_{max}^N-X_{0})^{Nk-1}}{\Gamma(Nk)\ (\tau_X/N)^{N k}} \exp \left( -\frac{\bar{X}_{max}^N-X_{0}}{\tau_X/N} \right) } &  \bar{X}_{max} \geq X_{0} \\
  0                                 &  \bar{X}_{max} < X_{0}
\end{array}  \right. \! \! \! \! .
\end{equation}

\begin{figure}[!h]
\centering
\includegraphics[width=7.5cm]{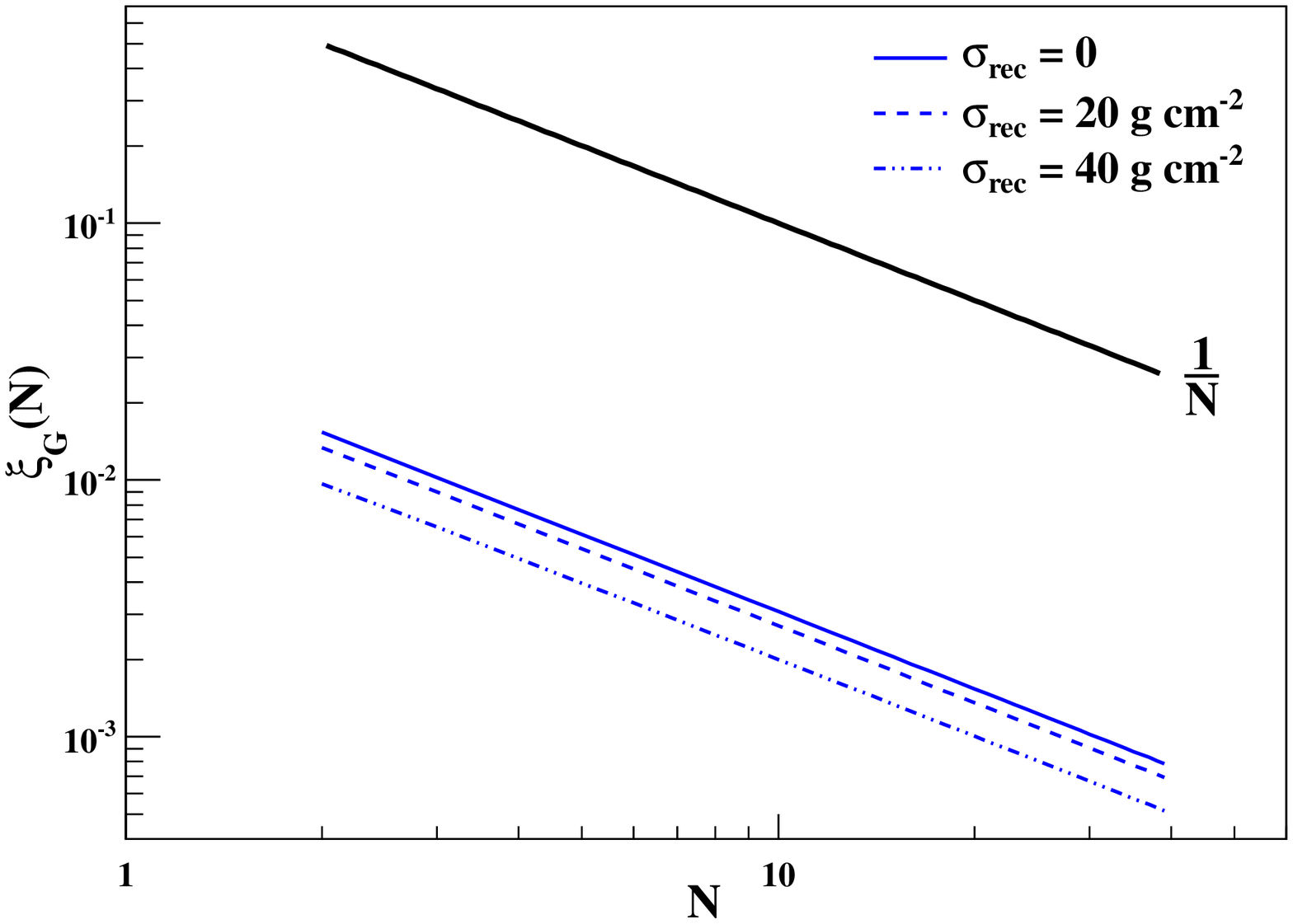}
\includegraphics[width=7.5cm]{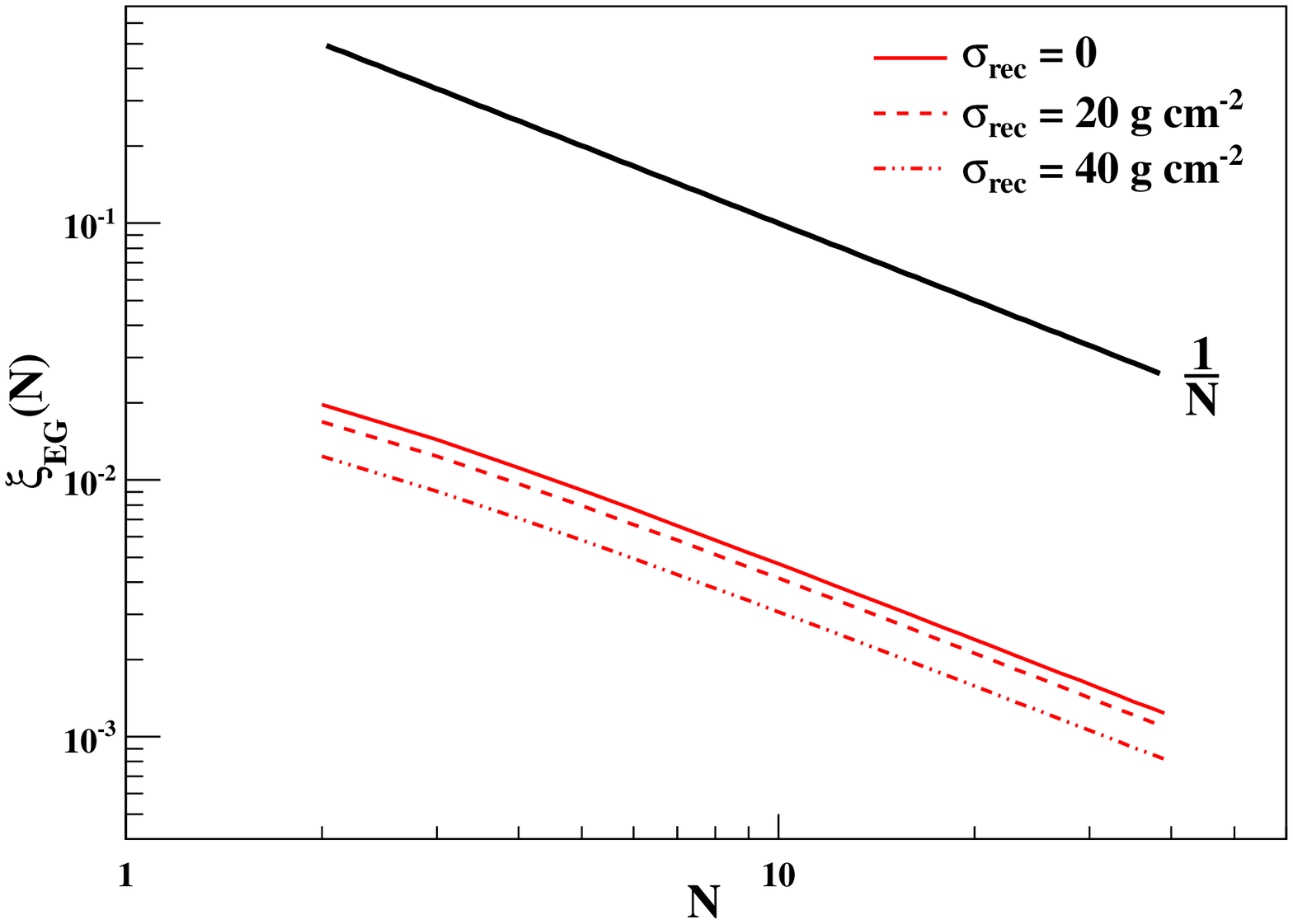}
\caption{$\xi$ as a function of the sample size $N$ corresponding to proton showers of $\log(E/eV)=19.5$, obtained 
for the shifted-Gamma distribution (left panel) and for the Exp-Gauss distribution (right panel). Blue and red solid 
lines correspond to the ideal case in which $X_{max}$ is determined without any uncertainty. Dashed and dashed-dotted 
lines correspond to the cases in which there is a Gaussian uncertainty on the determination of $X_{max}$ of 
$\sigma_{Rec}=20$ g cm$^{-2}$ and $\sigma_{Rec}=40$ g cm$^{-2}$, respectively. The hadronic interaction model used 
is QGSJET-II.}
\label{Bias}
\end{figure}

By using Eq. (\ref{AvGamma}) it is easy to show that,
\begin{equation}
\xi_G(N)=\frac{\sigma[X_{max}]}{\sqrt{k}\ \langle X_{max} \rangle}\ \frac{1}{N}.
\label{xiG}
\end{equation}
In this case, $\xi$ is also proportional to $1/N$ but it is suppressed by the ratio between the standard 
deviation and the mean value of $X_{max}$. A similar expression is obtained when the distribution function of 
$X_{max}$ is described by a truncated exponential function, see \ref{app2} for details. The blue solid line 
on the left panel of Fig. \ref{Bias} corresponds to $\xi_G$ as a function of $N$ for $\log(E/eV)=19.5$, 
approximately the mean value of the energy (weighted by the spectrum) for the last bin considered in Ref. 
\cite{Auger:10}. Note that, the number of events in this bin is 34. From the figure, it can be seen that 
$\xi_G$ is more than one order of magnitude smaller than the function $1/N$.

The distribution function of $X_{max}$ is affected by the presence of fluctuations introduced by the detectors. The 
distribution function of $\bar{X}_{max}^N$, including a Gaussian uncertainty on the determination of $X_{max}$ is given 
by,
\begin{equation}
\bar{P}_{G}^R(\bar{X}_{max}^N)=\frac{\sqrt{N}}{\sqrt{2\pi} \sigma_{Rec}} \int_0^\infty dX\ \bar{P}_{G}(X) %
\exp\left(-\frac{(\bar{X}_{max}^N-X)^2}{2 \sigma_{Rec}^2/N}    \right),
\label{PGavDX}
\end{equation}
where $\sigma_{Rec}$ is the standard deviation of such uncertainty. The mode of this distribution
is calculated numerically. Dashed and dashed-dotted lines on the left panel of Fig. \ref{Bias} correspond
to parameter $\xi_G(N)$ obtained for $\sigma_{Rec}=20$ g cm$^{-2}$ and $\sigma_{Rec}=40$ g cm$^{-2}$,
respectively. When a symmetric uncertainty on the determination of $X_{max}$ is included, the parameter $\xi$ 
becomes still smaller and decreases for increasing values of the uncertainty. This is due to the fact that
$\xi$ is larger for asymmetric distributions, like the exponential, and the convolution of the pure $X_{max}$ 
distribution with a Gaussian is more symmetric than the original one.    

The characteristic function of the Exp-Gauss distribution is the product of the characteristic function 
of the exponential distribution, $(1-i \lambda t)^{-1}$, with the one corresponding to a Gaussian, 
$\exp (i \alpha t -\beta^2 t^2/2)$. Then, the characteristic function of $\bar{X}_{max}^N$ is then given by,
\begin{equation}
\phi_{\bar{X}_{max}^N}^{EG}(t) = \left(1-i \frac{\lambda}{N} t \right)^{-N} \exp \left(i \alpha t -\frac{\beta^2}{N} t^2/2 \right),
\label{phiEG}
\end{equation}
which corresponds to the convolution of a Gamma distribution with a Gaussian,
\begin{eqnarray}
\bar{P}_{EG}(\bar{X}_{max}^N)&=& \frac{N^{N+1/2}}{\sqrt{2\pi} \beta \lambda^N \Gamma(N)} \int_{-\infty}^{\bar{X}_{max}^N} \! \! du\ %
(\bar{X}_{max}^N-u)^{N-1} \exp \! \left( \! -\frac{\bar{X}_{max}^N-u}{\lambda/N} \right) \nonumber \\
&& \times \exp \! \left( -\frac{(u-\alpha)^2}{2 \beta^2/N} \right).
\label{EGAv}
\end{eqnarray}
Last integral is calculated numerically in order to obtain the mode of the resultant distribution. The solid red 
line in the right panel of Fig. \ref{Bias} shows $\xi_{EG}$ as a function of the sample size for $\log(E/eV)=19.5$. 
Note that $\xi_{G}$ is smaller than $\xi_{EG}$, this is due to the more extended tail to larger values of the 
Exp-Gauss distribution compared with the corresponding one to the shifted-Gamma distribution. In any case, 
$\xi_{EG}$ is still about one order of magnitude smaller than $1/N$. As for the case of the Gamma distribution, 
dashed and dashed-dotted red lines correspond to $\sigma_{Rec}=20$ g cm$^{-2}$ and $\sigma_{Rec}=40$ g cm$^{-2}$, 
respectively. In this case the effect of the uncertainty on the determination of $X_{max}$ is included in 
$\bar{P}_{EG}$ just by replacing the parameter $\beta$ by $\tilde{\beta}=\sqrt{\beta^2+\sigma_{Rec}^2}$. As 
expected, the curves that include the uncertainty on the determination of $X_{max}$ fall bellow the one  
corresponding to the ideal case. 

\begin{figure}[!h]
\centering
\includegraphics[width=7.5cm]{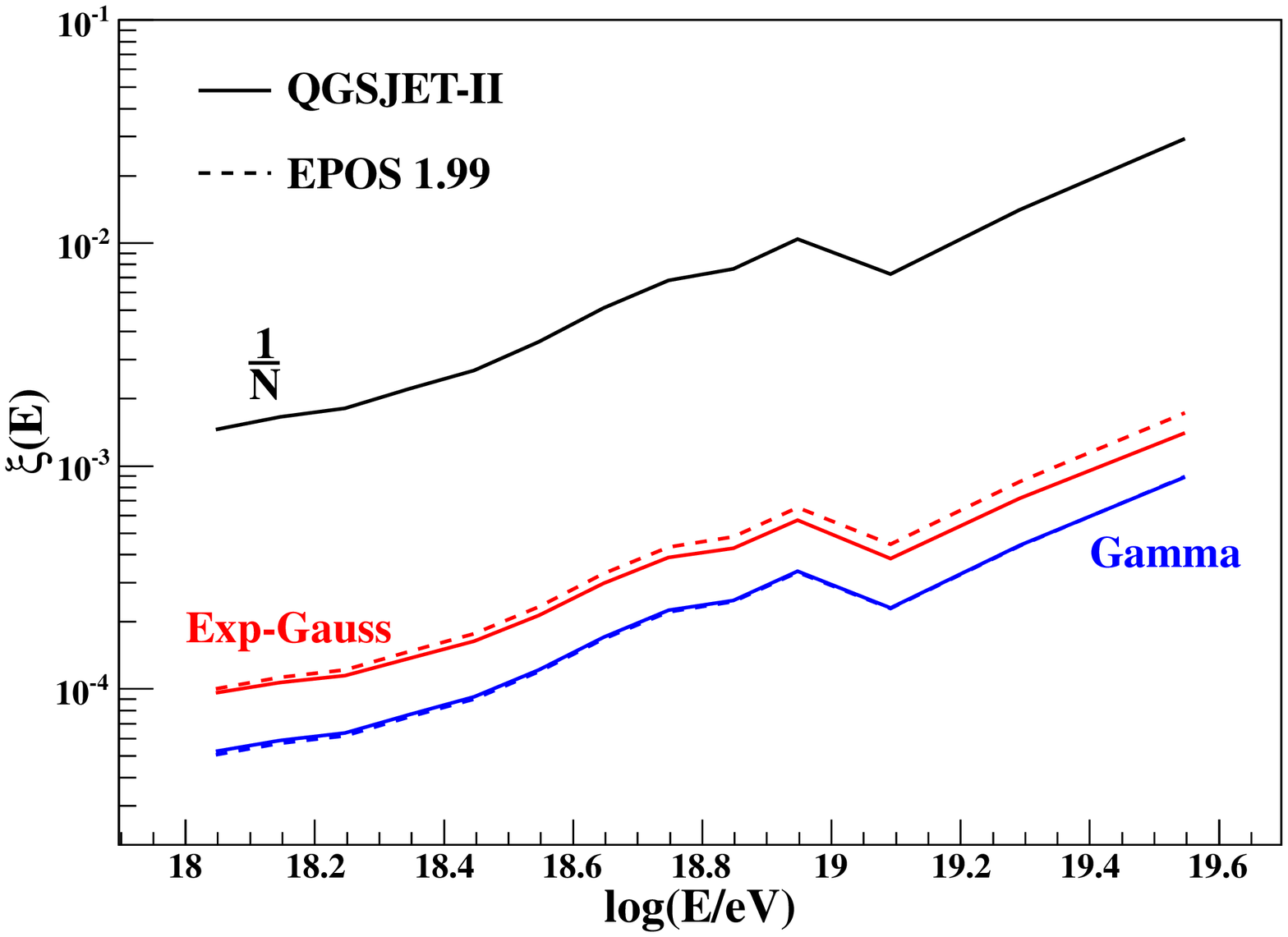}
\includegraphics[width=7.5cm]{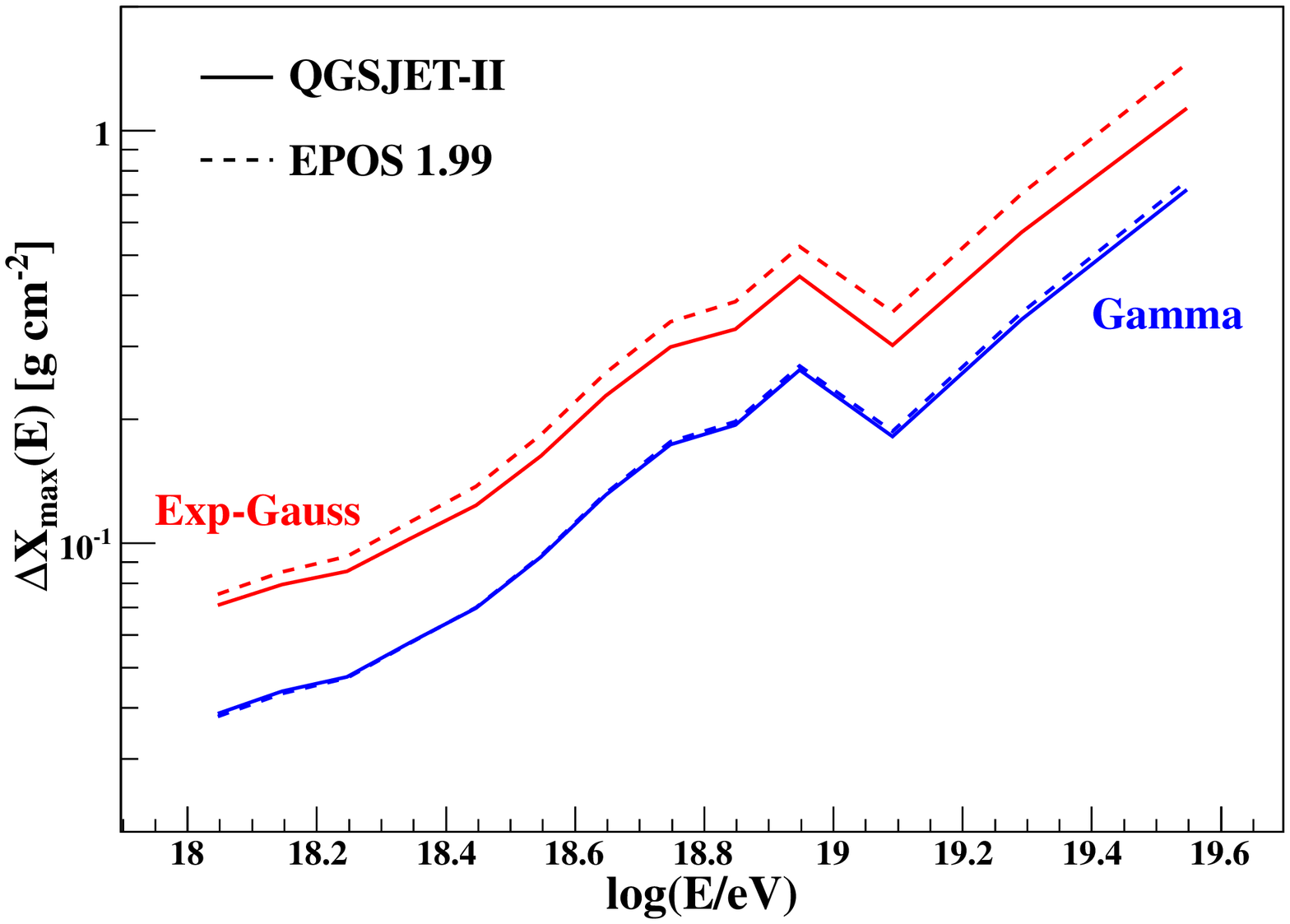}
\caption{$\xi$ (left panel) and $\Delta X_{max}$ (right panel) as a function of $\log(E/eV)$, for the statistics of 
the Auger data of Ref. \cite{Auger:10}. Solid lines correspond to QGSJET-II and dashed lines correspond EPOS 1.99.}
\label{BiasAugerPRL}
\end{figure}

The left panel of Fig. \ref{BiasAugerPRL} shows the parameter $\xi$ as a function of energy corresponding to the 
number of events in each energy bin taken from Ref. \cite{Auger:10}, for the case in which there is no uncertainty
on the determination of $X_{max}$ (which gives larger value of $\xi$, as shown before). The energy assigned to the 
$ith$ bin, used to calculate $\xi$, corresponds to the mean value of the energy in the bin weighted by the broken 
power law fit of the cosmic rays energy spectrum, $J(E)$, of Ref. \cite{AugerSpec:10},
\begin{equation}
\langle E_i \rangle = \mathop{\displaystyle \frac{\int_{E_i^L}^{E_i^U} dE\ E J(E)}{\int_{E_i^L}^{E_i^U} dE\ J(E)}},
\label{Ei}
\end{equation}
where $E_i^L$ and $E_i^U$ are the lower and upper limits of the $ith$ bin. It can be seen that the values of $\xi$, 
obtained by using the Exp-Gauss distribution and the shifted-Gamma distribution, are more than one order of magnitude 
smaller than the corresponding one for the exponential distribution, in the whole energy range and for both hadronic 
interaction models considered. As in the previous calculation, $\xi_{G}$ results are smaller than $\xi_{EG}$. In fact, 
the $\xi$ curve corresponding to the true distribution of $X_{max}$ should fall between the curves corresponding to 
the Exp-Gauss and the shifted-Gamma representation of the $X_{max}$ distribution. 

The right panel of Fig. \ref{BiasAugerPRL} shows the parameter $\Delta X_{max} = \langle X_{max} \rangle\ \xi$ which
gives the grammage of the shift suffered by $\langle X_{max} \rangle$ if $\bar{X}_{max}^N$ takes the value of the 
mode of its distribution. It can be seen, that for the last energy bin, the one with 34 events, $\Delta X_{max}$ 
is $\lesssim 1.5$ g cm$^{-2}$, which is much smaller than the systematic uncertainties on the determination of 
$\langle X_{max} \rangle$ estimated in Ref. \cite{Auger:10}. 

The energy bins considered in the analysis of Ref. \cite{Auger:10} have a width of $\Delta \log(E/\textrm{eV}) = 0.1$ in 
the energy range from $E=10^{18}$ eV to $E=10^{19}$ eV. Between $E=10^{19}$ eV and $E=10^{19.4}$ eV, $\Delta \log(E/\textrm{eV})$ 
changes to $0.2$ and the last bin corresponds to $E\geq10^{19.4}$ eV. Therefore, the number of events per bin decreases in the 
energy range from $E=10^{18}$ eV to $E=10^{19}$ eV, it increases from $96$ in the bin $[10^{18.9},10^{19}]$ eV to $138$ in the 
bin $[10^{19},10^{19.2}]$ eV and then, it decreases for the last two bins. This change in the bin width generates the structure 
around $E\cong10^{19.1}$ eV seen on the curves of Fig. \ref{BiasAugerPRL}.

Note that $\xi_{EG}$ calculated by using EPOS 1.99 is larger than the corresponding one for QGSJET-II, this
is due to the fact that the $X_{max}$ distributions obtained with EPOS 1.99 are more asymmetric (increase faster, 
coming from small values of $X_{max}$, and have a more extended tail) than the corresponding ones to QGSJET-II.   

Concerning iron showers, it can be seen that $\xi$ takes smaller values than the ones for protons. This is due to 
the large suppression of fluctuations in iron showers, the ratio of the standard deviation to the mean value 
of $X_{max}$ is smaller than for protons, producing smaller values of $\xi$ (see Eq. (\ref{xiG})). In particular, 
$\xi_G^{fe} = \xi_G^{pr}/K$ where $K$ increases from $\sim 2.3$ at $E=10^{18}$ eV to $\sim 2.4$ at $E=10^{19.5}$ eV
for QGSJET-II.

\section{Conclusions}

In this work we studied in detail statistical bias in the determination of the mean value of $X_{max}$, suggested 
in Ref. \cite{Wilk:11}, as a possible explanation of the deviation of Auger data from the proton expectation. 
We used two different functions to fit the $X_{max}$ distribution obtained from simulations: ($i$) the convolution 
of an Exponential distribution with a Gaussian and ($ii$) a shifted-Gamma distribution. We find that the bias 
obtained by using these two functions is more than one order of magnitude smaller than the corresponding one of 
the Exponential distribution, the one used in Ref. \cite{Wilk:11}. We find that the values of the bias, obtained 
for the convolution of the Exponential function with the Gaussian, are larger because it presents a more extended 
tail to larger values of $X_{max}$ than the shifted-Gamma distribution. We also find that the bias diminishes when 
a Gaussian (symmetric) uncertainty on the determination of $X_{max}$ is included. 

We also calculated the expected bias, as a function of primary energy, using the actual number of events in each 
energy bin of the Auger data, published in Ref. \cite{Auger:10}, for both hadronic interaction models considered in 
this work, QGSJET-II and EPOS 1.99. We find that the largest value of the bias, corresponding to the bin with 
the smallest number of events, is smaller than $1.5$ g cm$^{-2}$, much less than the systematic errors on the 
determination of $\langle X_{max} \rangle$ estimated in Ref. \cite{Auger:10}.     

\appendix

\section{Parameters for the Exp-Gauss fits}
\label{app1}

\begin{figure}[!h]
\centering
\includegraphics[width=7.5cm]{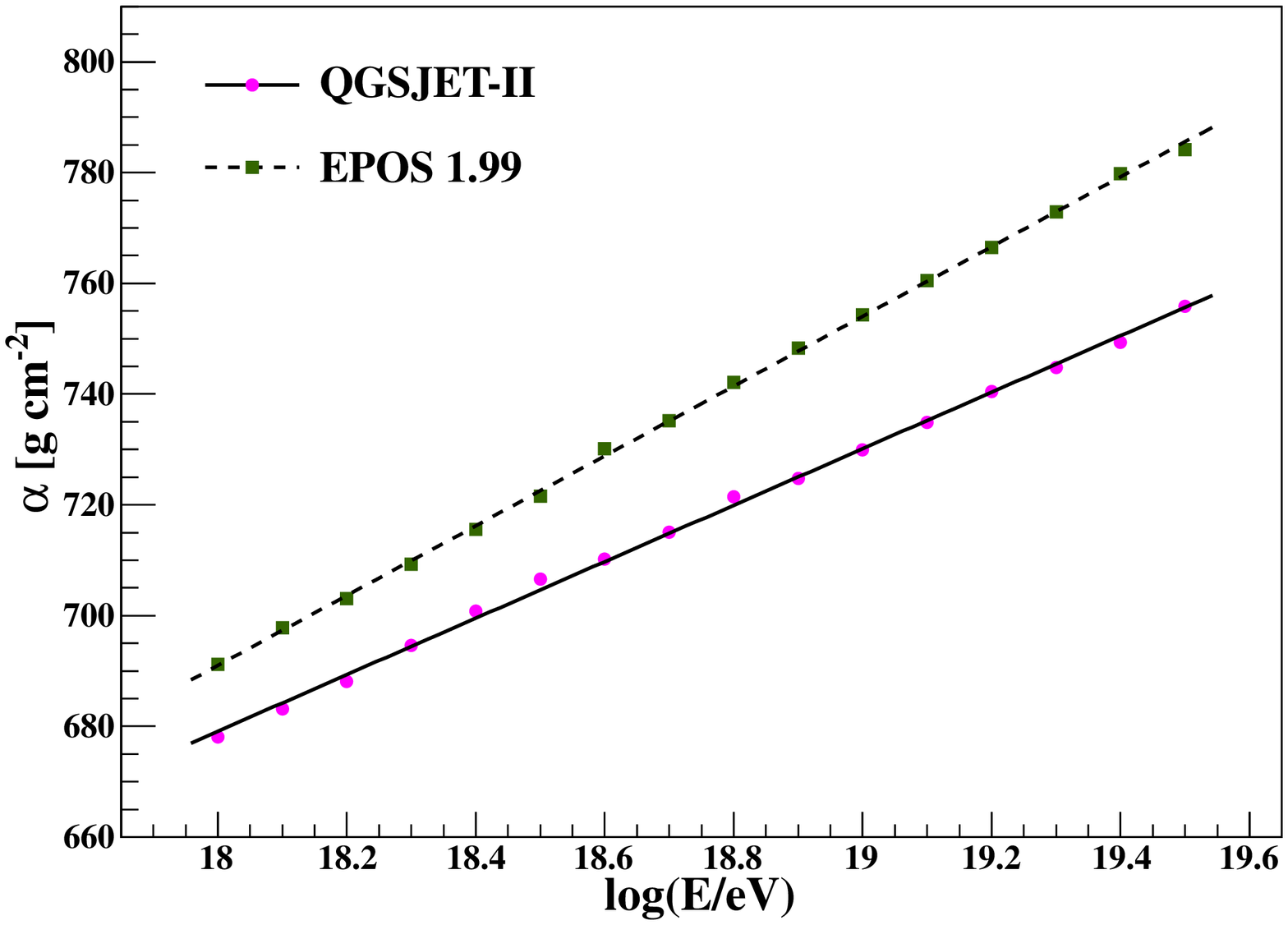}
\includegraphics[width=7.5cm]{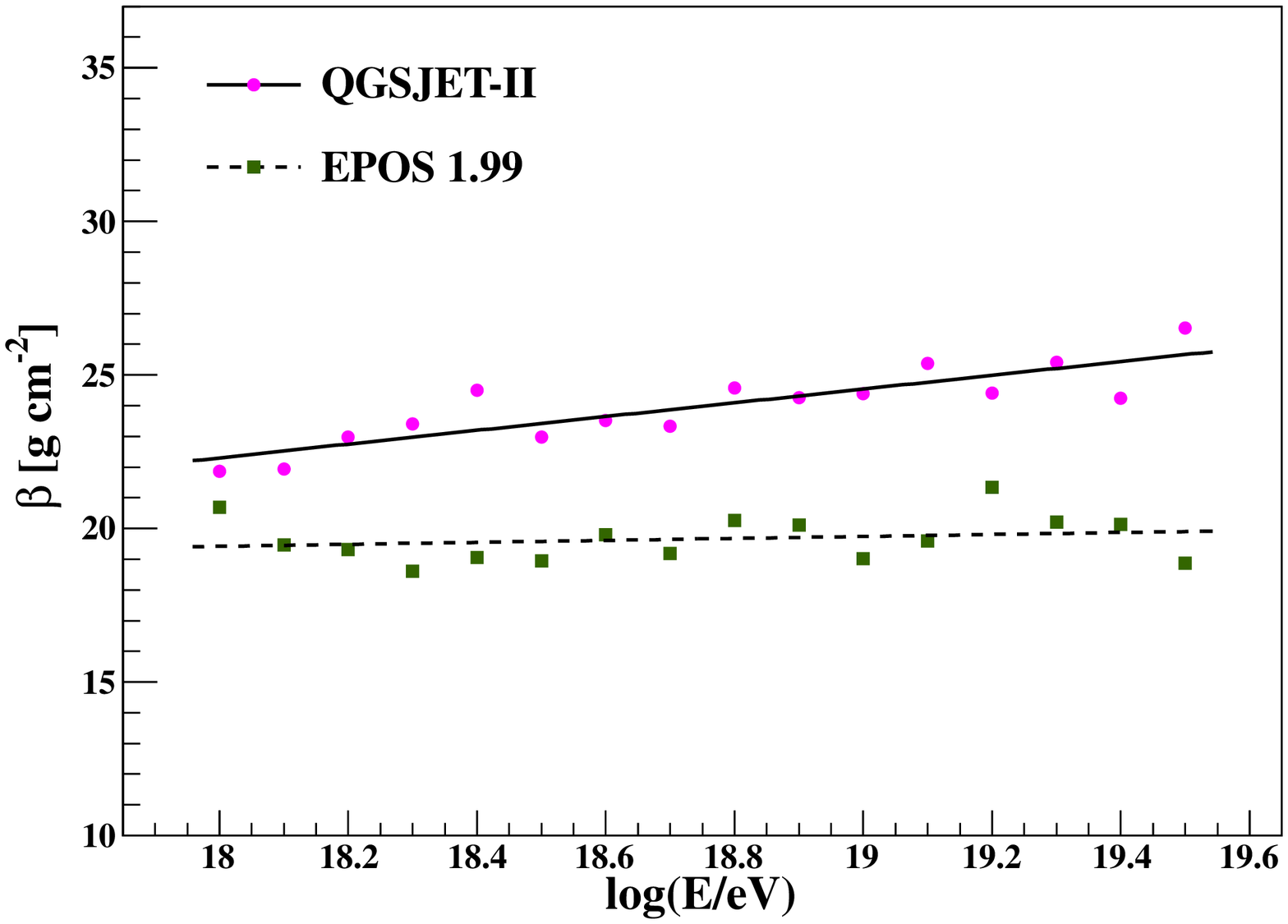}
\includegraphics[width=7.5cm]{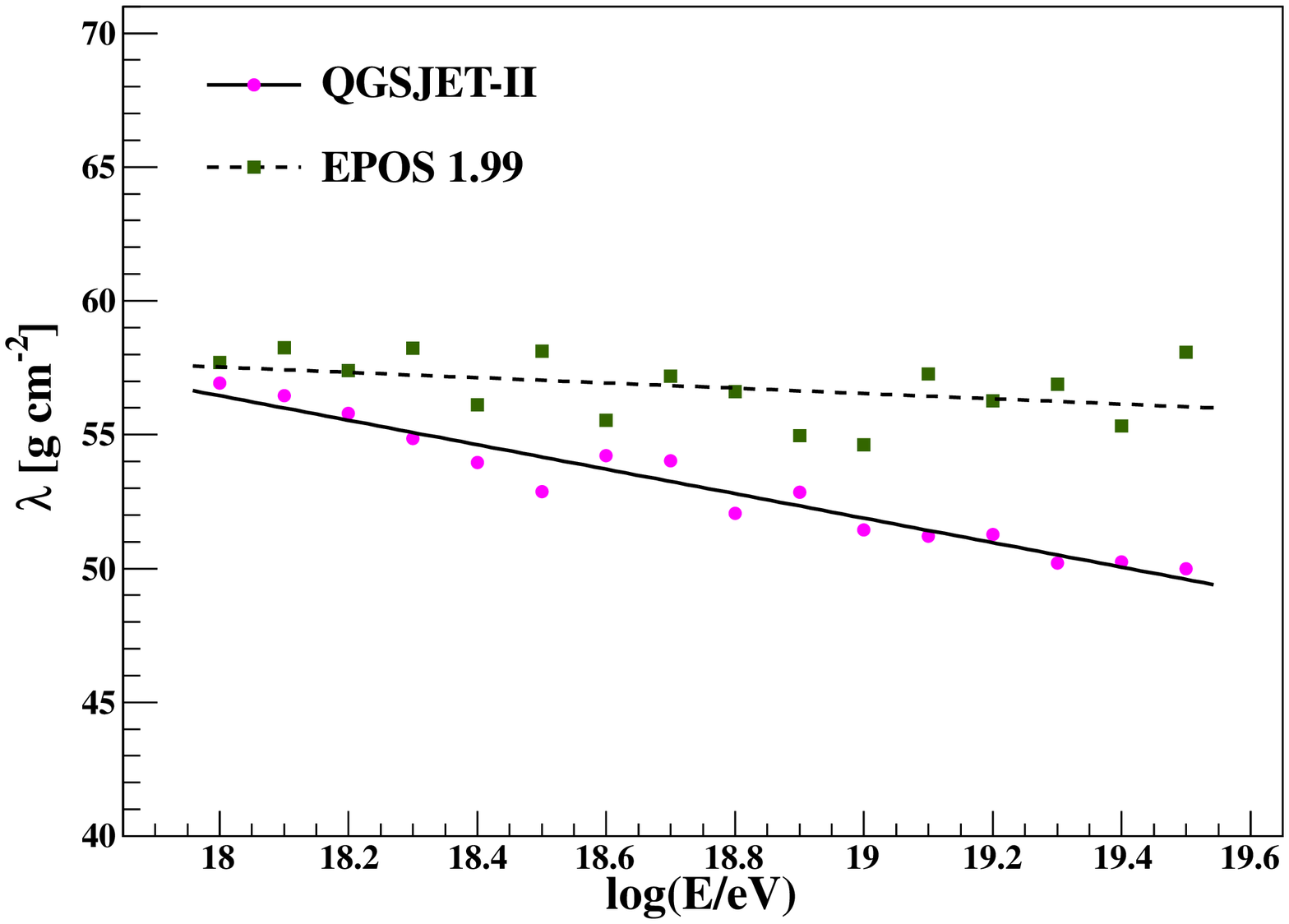}
\caption{Parameters $\alpha$, $\beta$ and $\lambda$ corresponding to the fits of the $X_{max}$ distribution with 
the Exp-Gauss function for QGSJET-II and EPOS 1.99. The straight lines correspond to the linear fits of the points.
\label{GEParam}}
\end{figure}

The parameters $\alpha$, $\beta$ and $\lambda$, obtained from the fits of the $X_{max}$ distributions with the Exp-Gauss 
function (see Eq. (\ref{ExpGauss})), are fitted with linear functions of $\log(E/eV)$ as shown in figure \ref{GEParam}. 
They can be written in the following way,  
\begin{equation}
\label{GEParamEq}
\left(
\begin{array}{c}
\alpha(E)  \\
\beta(E)   \\
\lambda(E) \\
\end{array} \right) =
\left(
\begin{array}{cc}
C_1 & C_2  \\
C_3 & C_4   \\
C_5 & C_6 \\
\end{array} \right) 
\left(
\begin{array}{c}
1           \\
\log(E/eV)  \\
\end{array} \right),
\end{equation}
where the coefficients $C_i$, $i={1...6}$, are given in table \ref{GEParamT} for both hadronic interaction models 
considered. 
\begin{table}[!h]
\caption{\label{GEParamT} Coefficients $C_i$, in [g cm$^{-2}$], corresponding to QGSJET-II
and EPOS 1.99.}
\begin{tabular*}{\textwidth}{@{}l*{15}{@{\extracolsep{0pt plus12pt}}l}}
\br
& $C_1$ & $C_2$ & $C_3$ & $C_4$ & $C_5$ & $C_6$ \\
\mr
QGSJET-II & -239.053 & 51.0096 & -18.0164 & 2.23981  & 138.806 & -4.57508  \\
EPOS 1.99 & -443.120 & 63.0078 & 13.6458  & 0.320823 & 75.2960  & -0.987241 \\
\br
\end{tabular*}
\end{table}

\section{Calculation of $\xi$ for a truncated exponential distribution}
\label{app2}

It is possible to describe the $X_{max}$ distribution function by a truncated exponential distribution,
which is given by,
\begin{equation}
\label{ExpT}
P_{TE}(X_{max})=\left\{ 
\begin{array}{ll}
  \mathop{\displaystyle \frac{1}{\Lambda} \exp \left( -\frac{X_{max}-X_{c}}{\Lambda} \right) } &  X_{max} \geq X_{c} \\
  0                                 &  X_{max} < X_{c}
\end{array}  \right.,
\end{equation}
where $\Lambda$ is a parameter that describe the tail of the $X_{max}$ distribution and $X_c$ is the truncation value.

The characteristic function of this distribution is, $\phi_{X_{max}}^{TE}(t) = \exp(i t X_c)\ (1-i t \Lambda)^{-1}$ and 
then, the characteristic function of the sample mean is given by, 
$\phi_{\bar{X}_{max}^N}^{TE}(t) = \exp(i t X_c)\ (1-i t \Lambda/N)^{-N}$, which corresponds to a shifted Gamma distribution.
Therefore, the distribution function of the sample mean is given by,
\begin{equation}
\label{ExpTSM}
\bar{P}_{TE}(\bar{X}_{max}^N)=\left\{ 
\begin{array}{ll}
  \mathop{\displaystyle \frac{(\bar{X}_{max}^N-X_c)^{N-1}}{\Gamma(N) (\Lambda/N)^N} \exp \! \left( \! -\frac{X_{max}-X_{c}}{\Lambda/N} \right) } &  \bar{X}_{max}^N \geq X_{c} \\
  0                                 &  \bar{X}_{max}^N < X_{c}
\end{array}  \right. \! \! \! \! \! .
\end{equation}

By using that $\langle X_{max} \rangle = X_c + \Lambda$, it is easy to show that,
\begin{eqnarray}
\label{XiTEa}
\xi_{TE}(N) &=& \frac{\Lambda}{\Lambda+X_c}\ \frac{1}{N}, \\
\label{XiTEb}
&=& \frac{\sigma[X_{max}]}{\langle X_{max} \rangle}\ \frac{1}{N}.
\end{eqnarray}
Note that, it can be seen, form Eq. (\ref{XiTEb}), that $\xi_{TE}$ takes a very similar form to the one obtained 
for the shifted-Gamma function, see Eq. (\ref{xiG}).

Typical values of the parameters, obtained experimentally, are $X_c\cong700$ g cm$^{-2}$ and $\Lambda \cong 56$ 
g cm$^{-2}$ \cite{Ulrich:11} (note that these parameters depend on primary energy and the ones used here, obtained 
from Ref. \cite{Ulrich:11}, correspond to the energy interval $[10^{18}, 10^{18.5}]$ eV, in any case, they are just 
used to roughly estimate the suppression factor of the bias). Therefore, $\xi_{TE}(N) \cong 0.125/N$, which is 
suppressed by a factor $0.125$ with respect to the corresponding one to the exponential distribution.

\end{document}